\documentclass{aa}

\usepackage{graphicx}
\newcommand{\filter}[1]{\mbox{\it #1\/}}              

\usepackage{txfonts}
\usepackage{xcolor}

\begin{document}

   \title{LZ-STAR Survey: Low-metallicity Star Formation Survey of Sh2-284. II. The initial mass function}

   \author{M. Andersen
          \inst{1}
          \and
          A. Brizawasi \inst{2,3}
            \and
          Y. Cheng\inst{4}
          \and 
          R. Fedriani \inst{5} 
          \and 
          J. J. Armstrong \inst{3}
          \and 
          M. Robberto \inst{6} 
          \and 
          M. Aghakhanloo \inst{7}
          \and 
          J.~C.~Tan \inst{3,7} 
                    }

   \institute{European Southern Observatory, Karl-Schwarzschild-Strasse 2, D-85748 Garching bei München, Germany 
              \email{morten.andersen@eso.org}
            \and 
             Indian Institute of Science Education and Research (IISER), Mohali, India 
             \and            
                         Dept. of Space, Earth \& Environment, Chalmers University of Technology, SE-45293 Gothenburg,
Sweden 
\and  
            National Astronomical Observatory of Japan, 2-21-1 Osawa, Mitaka, Tokyo 181-8588, Japan 
             \and 
              Instituto de Astrof\'isica de Andaluc\'ia, CSIC, Glorieta de la Astronom\'ia s/n, E-18008 Granada, Spain 
             \and 
             Space Telescope Science Institute, Baltimore, MD, USA
             \and 
Dept. of Astronomy, Univ. of Virginia, Charlottesville, VA 22904, USA
                     }

   \date{}
 
  \abstract
   {   To fully understand the star formation process, we are compelled to study it in a variety of environments. 
   Of particular interest are how star formation and the resulting initial mass function (IMF) vary as a function of metallicity.} 
   {Using JWST/NIRCam, we have observed an embedded young cluster in the vicinity of Sh2-284 (hereafter S284), the HII region associated with the open cluster Dolidze 25, with the aim to study star formation in a metal-poor, i.e., about 1/3 of solar, environment. In particular, we aim to measure the peak of the IMF.}  
  {Using JWST/NIRCam photometry, we identified the embedded cluster S284-EC1 and resolved its low-mass content. By comparison with pre-main sequence evolutionary tracks we determine the mass and extinction for the individual cluster members. Extinction limited samples are created based on the distribution of extinction and the completeness of the data. { For the region with a completeness of 60\%\ or higher, 
   we have fitted a log-normal distribution to the IMF.}}
   { 
   { 
    Adopting an age of 1~Myr of the cluster members, supported by the high extinction and outflow activity in the region,  the peak of the IMF is at $m_c=0.17\pm0.02 M_\odot$. {  This value lies toward the low-mass end of the range reported for local young clusters in the literature} 
    However, the result is sensitive to the age of the cluster. If an older age of 2 Myr is adopted, the peak mass increases to $0.28\:M_\odot$.}
   }
   {We have found evidence for IMF variation as a function of metallicity, i.e., the peak of the IMF shifts to lower masses as one goes from solar to 1/3 solar metallicity. However, we caution that the result is sensitive to the assumed age of the stellar population, i.e., with peak mass rising if an age older than 1~Myr is adopted. This study further motivates the need for expanded samples of low-metallicity regions and their content to enable more comprehensive measures of the IMF in such environments.
   }


\maketitle

\section{Introduction}
One of the fundamental outcomes of the star formation process is the distribution of masses of stars and brown dwarfs, i.e., the initial mass function (IMF).  A key question is  whether star formation, including the resulting IMF, depends on global environmental parameters, such as metallicity, density and magnetic field strength. 
Substantial work has been carried out to obtain measures of the IMF of the local Galactic field population and in relatively nearby star-forming regions. The main conclusions are that there is little evidence for variations in the observed stellar and brown dwarf IMF down to about $0.03\:M_\odot$ ($30 M_\mathrm{Jup}$) \citep{andersen_08,hennebelle,kroupa2024}.  For a log-normal representation, the field and nearby star-forming regions are consistent with an IMF with a peak at $0.25\:M_\odot$ \citep{chabrier05}. 

Some star-forming theories predict variations in the IMF as a function of environment. 
In particular, the effect of metallicity has been discussed in the literature, but with a variety of different conclusions. 
Using hydrodynamic simulations of star-forming clouds with metallicities from $Z=10^{-1}$ to $10^{-4}Z_\odot$, \citet{chon2021} found the peak of the IMF  declines with decreasing metallicity. Their case with $Z=10^{-1}Z_\odot$ yielded a median mass, i.e., close to the peak value, of $0.14\:M_\odot$. However, these simulations did not include any effects of magnetic fields or stellar feedback.
On the other hand, \citet{Guszejnov2022} found the opposite effect. In contrast, \citet{myers2011} and \citet{bate14} found little variation as a function of metallicity. Thus, on the theoretical side, the answer is unclear, likely because of the incomplete treatment of the full physics of star formation and feedback. 

In any case, different theoretical predictions need to be tested with observations of the IMF, especially targeting the value of its peak mass. However, nearby regions only cover a small range in 
metallicity, i.e., with values comparable to solar. Thus it is not practical to use these regions to examine dependence of the IMF on metallicity. 
Studies of massive, more distant Galactic clusters have suggested the IMF to be consistent with the local field IMF down to $0.2\:M_\odot$ \citep{andersen_wd1}, but these systems also have near solar metallicity. Similar studies in the large Magellanic Cloud (LMC), where the metallicity is about half solar, have probed the IMF down to about $1\:M_\odot$ \citep{andersen_30dor}, also finding general consistency with the local field IMF. A similar conclusion was reached for the lower metallicity cluster NGC 602 in the small Magellanic Cloud (SMC) \citep{schmalzl}. 
Similarly, the field IMF in the SMC has been determined down to $0.37\:M_\odot$ without any evidence for an IMF varying with metallicity \citep{kalirai}. 
Even lower metallicities ($[\mathrm{Fe/H]=-2.6}$) have been probed in the Coma Berenices Dwarf Galaxy, where the IMF could be traced down to $0.23\:M_\odot$ \citep{gennaro_dwarf}, with the conclusion that the derived IMF was also consistent with that of the Galactic field. 

Common for the low-metallicity studies quoted above is the fact that they barely reach the Galactic field IMF peak mass of $0.25\:M_\odot$ \citep{chabrier05}. 
Although nearby galaxies are prime targets for studies of the IMF in metal-poor conditions, such studies are restricted by their large distances of 50-60~kpc for the Magellanic Clouds and 43~kpc for Coma Berenices. This problem can be overcome by studying the low-mass IMF in metal-poor environments in the outer Galaxy, given its global metallicity gradient \citep{daflon}. \citet{yasui23} studied the IMF in the Galactic star-forming region Sh2-209 region, which has an oxygen abundance of $-0.5$~dex compared to solar.
They derived a relatively shallow, i.e., top-heavy, high-end power law index of $\Gamma=-1.0$ (c.f., the Salpeter IMF value of $-1.35$), but with a break mass of $0.1\:M_\odot$ that is smaller than derived from equivalent fitting in the solar neighborhood, i.e., $\sim0.3\:M_\odot$.
\citet{zinnkann} expanded on these results 
to reach similar conclusion that a denser cluster at low metallicity leads to a top-heavy IMF. 

Although recent James Webb space telescope (JWST) observations have allowed the detection of objects down to below a Jupiter mass in nearby star-forming regions \citep[e.g.,][]{defurio}, identification of pre-main sequence stars in the distant local group galaxy Wolf–Lundmark–Melotte  \citep{kalari25} and objects down to or even below the brown dwarf limit in the Magellanic Clouds \citep{zeidler24}, the lowest masses are still most easily reached in the few low-metallicity Galactic young star clusters that are known. The first steps in this direction using JWST were by \citet{yasui24} of the Digel 2 cloud where a peak mass of the IMF of  $\sim0.03\:M_\odot$ (with about a factor of 3 uncertainty) was found, substantially below the field IMF peak value of $0.25\:M_\odot$. 
However, the relatively large uncertainty associated with this measurement indicates that a wider variety of young clusters need to be studied to more securely probe the low-mass IMF in these low-metallicity environments.

The Sh2-284 (hereafter S284) HII region, associated with the  Dolidze 25 cluster, is a prime test bed to determine the low-mass IMF in a metal-poor environment. It is located in the Galactic anti-center directions at ($l,b$)=(211\fdg9, -01\fdg3), and has been found to have a $\sim1/3-1/2$ solar metallicity \citep{neugerella15}. 
The distance to the cluster has been under debate, suggested to be at 5.5 kpc \citep{lennon90} to 6 kpc \citep{russeil2007}. 
However, later determination using more recent low-metallicity stellar evolutionary models have suggested a  distance of, 4 kpc \citep{cusano2011}, and $4.5\pm0.3\:$kpc \citep{neugerella15}. 
\citet{neugerella15} noted that 4.5 kpc is in agreement with a maser 2 degrees away having a similar radial velocity as the stars in Dolize 25 and thus likely at the same distance. 
\citet{guarcello} used {\it GAIA}  early data release 3 to determine a distance of 4.5$\pm$0.5kpc for the spectroscopically verified members of Dolidze 25 in \citet{neugerella15} that were not identified as binaries. 
We have repeated the analysis with the current {\it GAIA 3} catalogue \citep{gaia} , taking into account the individual parallax zero-points not available in the previous study. { Using { GAIA} stars that are confirmed members of Dolidze 25 by spectroscopy and not a spectroscopic binary in \citet{neugerella15}, with a RUWE parameter value $<1.4$ (five stars\footnote{The {\it GAIA} IDs are:  3113525799800744320, \\ 3125505670478616192, 3125505704838338816, \\ 3125506014076025088, and 3125506151511957120}
}), we determine a distance of $4.2\pm0.7\:$kpc after correcting for the zero-point as prescribed in \citet{lindegren} and adopting 1 over the parallax as the distance. 
This is consistent with the previous {\it GAIA} estimate and also with the maser 2 degrees away. 
\citet{ashraf} analyzed {\bf GAIA}  DR3 data   for a 3\arcmin\ radius of the center of Dolidze 25 using DBSCAN to select members and determined a distance of $4.2\pm0.3\:$kpc, consistent with the value from spectroscopically selected members.  
Given these results,  we adopt the same distance of 4.5~kpc as in \citet{neugerella15} as a fiducial value, but explore the consequences of varying the distance by $\pm0.3\:$kpc.

  The age of the Dolidze 25 cluster has been estimated in multiple studies. With the distance of 4.5 kpc, \citet{neugerella15} estimated the age to be no more than 3 Myr based on the position in the color-magnitude diagram of the earliest-type cluster stars.  
\citet{guarcello} utilized an X-ray selected sample, using the 17\arcmin\  square field of view of Chandra ACIS-I, to determine the peak of the age distribution to be at 1 Myr, and the median at 1.2 Myr.  
\citet{kalari15} suggested a median age of 3.5~Myr based on optical spectroscopy of intermediate-mass stars around the center of the cluster. 
In contrast, \citet{ashraf} used MUSE spectroscopy of the central 1\arcmin$\times$2\arcmin\ (1.3~pc$\times2.6$~pc) of Dolidze 25 to derive an age distribution peaked at 1.53 Myr. The lower age compared to e.g. \citet{kalari15} was suggested to be based on a more comprehensive simultaneous fitting. They used simultaneous fitting of spectral type, accretion and veiling using low metallicity atmospheric models. 
\citet{patra24} used broad-band near-infrared imaging to investigate  three regions of 1-2\arcmin\ radius within the cluster (marked in Fig.~\ref{overview}). Their results were similar to those of \citet{guarcello,ashraf} with ages of 1.3-2 Myr for the sub-clusters and similar relatively modest extinction towards the sources of $A_V\sim3$. 

Dolidze 25 is surrounded by its HII region with an inner radius  of 15\arcmin , corresponding to 20 pc. Outside this radius, the HII region interacts with the surrounding molecular material evident at mid-infrared wavelengths due to the emission from the associated dust  (Fig. 1). 
A large scale mid-infrared survey of S284 was performed by \citet{puga09} using Spitzer IRAC imaging. 
Dolidze 25 is surrounded by molecular material where high concentrations of Class I  sources were detected. They interpret this to be consistent 
The age of Class I sources is in the range of 0.1-1 Myr, suggesting some age progression across the region, with the younger sources associated with higher extinction regions at the edge of the HII region.

\citet{puga09} identified their region RE 20 pc to the East of the center of Dolidze 25 as a region rich in  young sources. 
Subsequent far-infrared Herschel showed strong emission within region Re \citep{cheng_jwst} with a total gas mass of 760 M$_\odot$. The  over-density with a full width at half maximum of 0.65 px is hosting a class I star driving a molecular outflow discussed in \citet{cheng_jwst}. 

The current study focuses on the low-mass stellar and brown dwarf content of this embedded region which we name sh284-EC1. 
Due to its compactness, youth, and low metallicity, region RE is ideal to probe the early stages of star formation in a low-metallicity environment. 

As part of the Low-Metallicity Star Formation Survey in Sh2-284 (LZ-STAR), we present photometry using JWST/NIRCam observations of an embedded region in the shell of S284 within the RE region from \citet{puga09}, which we term S284 embedded cluster 1 (S284-EC1). Our goal is to assess the distribution of young objects and determine the IMF well into the sub-stellar regime. An overview of the region was presented by \citet{cheng_jwst} (LZ-STAR Paper I), including a census of the dense ALMA-detected gas cores and characterization of a potential massive protostar that appears to be forming near the center of the protocluster.

Our paper is structured as follows. In Section 2 we discuss the data, their reduction and the photometry. Section 3 presents the results from the photometry, i.e., the luminosity functions, the color-magnitude diagrams and the morphology of the cluster. Section 4 discusses the construction of the isochrones, the mass estimates for individual objects and the derived IMFs under different assumptions on metallicity and age. 
Finally, we conclude in Section 5. 

\section{Observations and data reduction}
\subsection{JWST data} 

\begin{figure}
   \centering
     \includegraphics[angle=0,width=9cm]{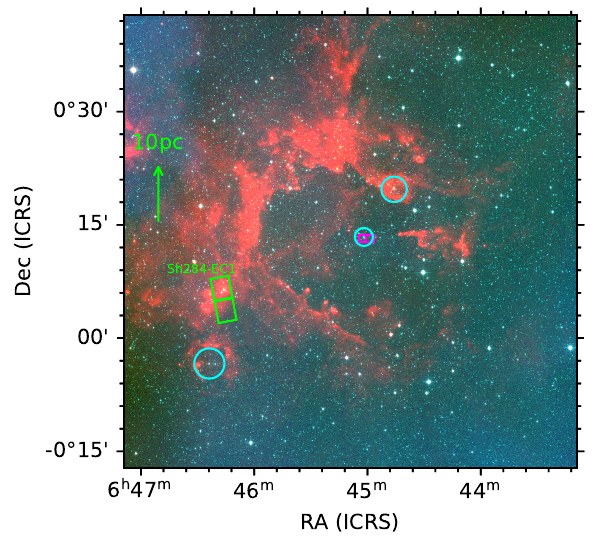}
 
      \caption{{ A one degree square field of view WISE 3.6 $\mu$m (red), 2MASS Ks, and H (green and blue) image composite of S284, centered on Dolidze 25 and with the target of this study, sh284-EC1 to the East-South-East. 
      A 10 pc scale bar is shown for a distance of 4.5kpc. sh284-EC1 is located 20pc to the East-South-East and the JWST/NIRCam pointing is overlayed as the green boxes with the A module to the North.           
       The fields in \citet{ashraf} and \citet{patra24} are shown as the magenta rectangle and the cyan circles, respectively.}}
         \label{overview_ls}
   \end{figure}

JWST/NIRCam observations were obtained on 19th Oct. 2022 of S284 through the \filter{F162M}, \filter{F182M}, \filter{F200W}, \filter{F356W}, \filter{F405N}, and \filter{F470N} bands in program ID 2317 (PI: Y. Cheng). The observed $2\arcmin \times 4\arcmin$ field is shown in Fig.~\ref{overview_ls}. It overlaps with the region 21 within the RE area in \citet{puga09} and was chosen due to the embedded sources identified in the Spitzer survey. 
  The three short wavelength filters of the observations were used to measure the stellar atmospheres of the individual sources. 
 Although a wider wavelength coverage would provide better estimates of extinction, it would create larger differences in the point-spread-functions, the sensitivity and crowding characteristics.
 
A total of six dithers and six groups resulted in a total integration time of 1739.357 seconds for the shallow2, and 644.206 seconds for the bright2 observations. The readout mode was shallow2 for \filter{F162M}, \filter{F182M}, \filter{F405N}, \filter{F470N} and bright2 for \filter{F200W} and \filter{F356W}. The dither type was fullbox and the pattern tight3, resulting in relatively homogeneous depth in the mosaic at the cost of a few small gaps in the final mosaics. 
 
The data were reduced using the JWST Calibration Pipeline Version=1.12.5. A few changes were employed to the default parameters. Through the stage 1 processing we allowed the use of the first read sampling up the ramp to increase the dynamic range of the data, limiting the number of saturated sources. Further, the frames were drizzled together using a final pixel scale of 0.03\arcsec, using the relative offsets determined by tweakreg. The data were combined independently for the A and B modules due to a small offset between the two in the mosaicing process. 

\begin{figure*}
   \centering
     \includegraphics[angle=0,width=18cm]{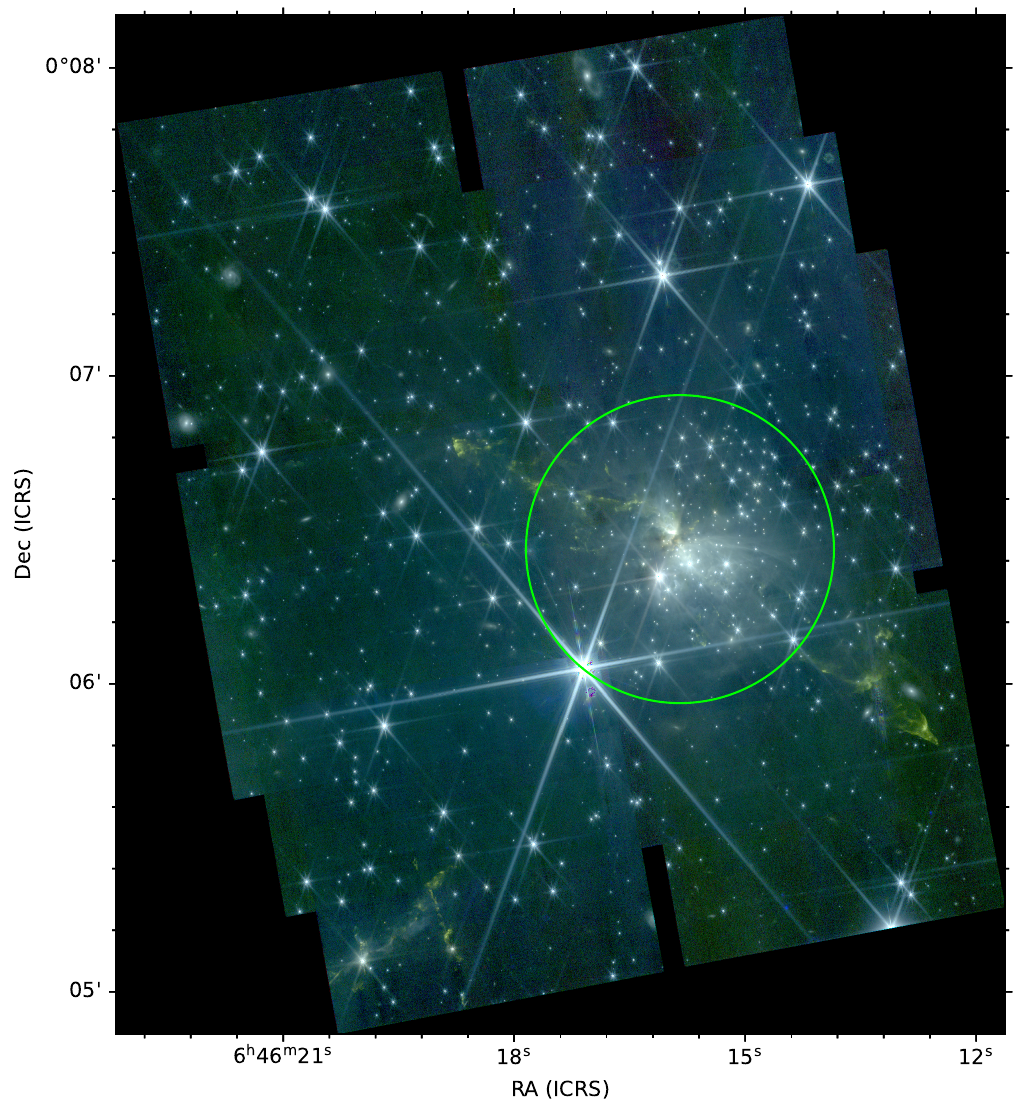}
 
      \caption{Color-composite of the short wavelength NIRCam A-module. Red is \filter{F200W}, green \filter{F182M}, and blue is \filter{F162M}, respectively. The circle indicates a 1000 pixel radius (0.5\arcmin, 0.65~pc), with the radius determined in Section \ref{cluster_specs}. The scale for all three color channels is logarithmic to emphasize both the bright and faint features.                  }
         \label{overview}
   \end{figure*}

Figure~\ref{overview} shows the three-color image of module A, using the three short wavelength bands. A relatively rich cluster is seen centered at (RA,DEC) = (6:46:15.8501,+0:06:26.197). The green circle with a radius of 1000 pixels (0.65~pc for a distance of 4.5 kpc) illustrates the radius where the cluster population surface number density is similar to the field (see Section 3). 

\subsection{Photometry}

Source detection within the cluster is complicated both by the strong nebulosity and the large dynamic range in the observations. Spurious detections are an issue because of the almost diffraction limited point-spread-function (PSF) that introduces artifacts from bright sources that can be identified as point sources. To identify and remove spurious detections we use an approach similar to \citet{andersen_wd1} where the multicolor data and the dependence on wavelength are utilized. Briefly, a spatially constant psf was created from the A module using twenty bright, non-saturated isolated stars across the field and a radius of the psf of 11 pixels adopting the {\tt iraf} implementation of {\tt daophot} \citep{stetson}. The brightest point source objects are identified in the \filter{F200W} image and removed using {\tt allstar}. A fitting radius of 3 pixels was used for all three bands. In the residual image the second brightest stars, at half the threshold as the previous step, are detected and added to the list of the brightest stars. This list is then used to remove stars from the original frame. This process is repeated with subsequently lower thresholds down to four times the background noise. The positions for all objects are then injected into {\tt phot} and {\tt allstar} again in the \filter{F200W} frame and psf photometry on all sources is performed in one step. The final source list for the \filter{F200W} is then used as input for the \filter{F162M} and \filter{F182M} filters, where PSF photometry is performed in a similar manner as for the \filter{F200W} band. The centering in each filter is then compared and a source is considered a real source (and not, e.g., a diffraction spot) if the spatial position agrees to within 0.5 pixels between the three bands. The list of vetted sources is then used for a final process through {\tt phot} and {\tt allstar} in each filter, ensuring only detections in all three filters are included.

The photometry in each band is zero point calibrated using the source list from the pipeline for the sources. Instead of converting the JWST photometry into a more standard system, e.g., the 2MASS system, we will use the models and isochrones converted into the JWST system (see below). 

\begin{figure}
   \centering
   \includegraphics[angle=0,width=9cm]{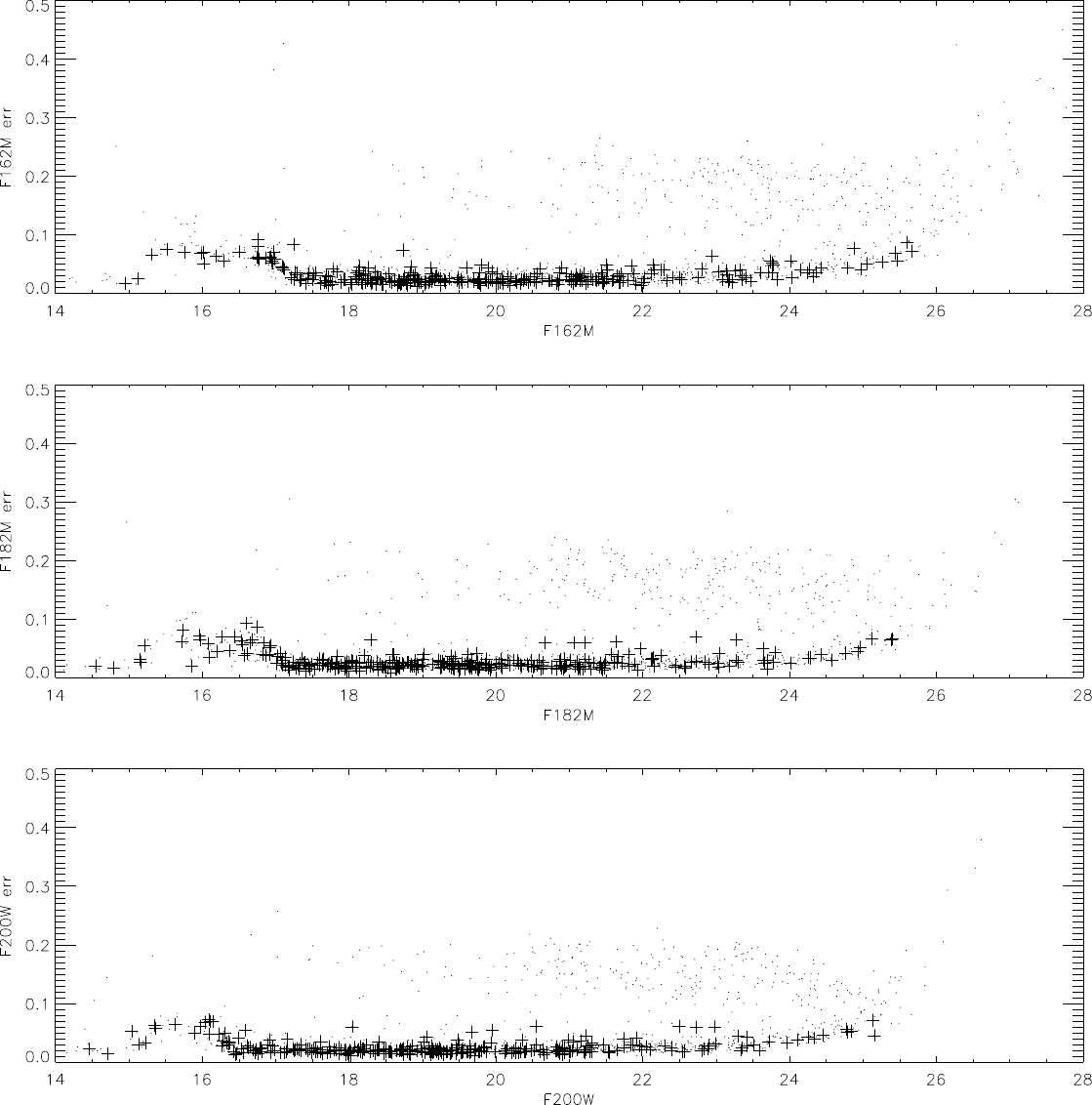}
      \caption{The photometric errors as a function of magnitude for the stars detected in all the \filter{F162M}, \filter{F182M}, and \filter{F200W} bands in the A module. The thicker cross symbols are for the objects within the designated cluster radius of 1000 pixels.               }
         \label{mag_errors}
   \end{figure}

The magnitude errors as a function of magnitude for each filter in the A module are shown in Fig.~\ref{mag_errors}. The sources within the central 1000 pixels of the cluster center, defined below, are marked with thick symbols. The magnitude error generally increases at fainter magnitudes, as expected, but an increase is also seen at bright magnitudes for all three filters. This is due to the usage of the first readout where the zero point for the detector is poorly determined and thus for bright stars the point-spread function will deviate in the core from the model PSF. The mass limits for the extinction limited samples do not include these objects. 

For all magnitudes there is a presence of relatively high photometric error points, above 0.1 mag. Upon visual inspection it is clear that these are due to detections of slightly extended objects, typically galaxies and nebulosity structures, which are then excluded due to the photometric error cuts. These objects further had poor reduced chi square fits, due to their extended nature.  The total number of objects detected with photometric errors smaller than 0.1 mag in  all three bands is 1092 for the A module and 874 for the B module, respectively.
 
\subsection{Completeness corrections}

The sensitivity of the observations is complicated to estimate due to the relatively high source density and nebulosity within the cluster region and these factors are potential sources of biases in the source catalog. Artificial star experiments have been carried out to quantify how source counts might have been affected as a function of magnitude and as a function of position in the field of view. Following the approach in, \citep[e.g.,][]{andersen_30dor,gennaro,andersen_wd1}, we have inserted point sources with colors and magnitudes similar to the observed objects. To avoid affecting the characteristics of the observations, 25\%\ of the detected sources have been inserted, i.e., 300-350 sources per run. This is a slightly higher fraction than in, e.g., \citet{andersen_wd1}, which is allowed by the level of crowding being relatively low in S284-EC1.  Indeed, the main effects for this cluster are the nebulosity and diffraction spikes from the bright stars. Further, the relative surface density profile of the inserted sources is similar to that of the observed objects. This was in practice done by for each object selected as an artificial star from the source list, the distance from the assigned cluster center was calculated and the artificial source was then located at the same distance, but at a random angle. A total of 2000 runs with different random star populations were performed. The data analysis was identical as for the science data and an artificial star was considered retrieved if the recovered source was within 0.5 pixels of the true position. 

\begin{figure}
   \centering
   \includegraphics[angle=0,width=9cm]{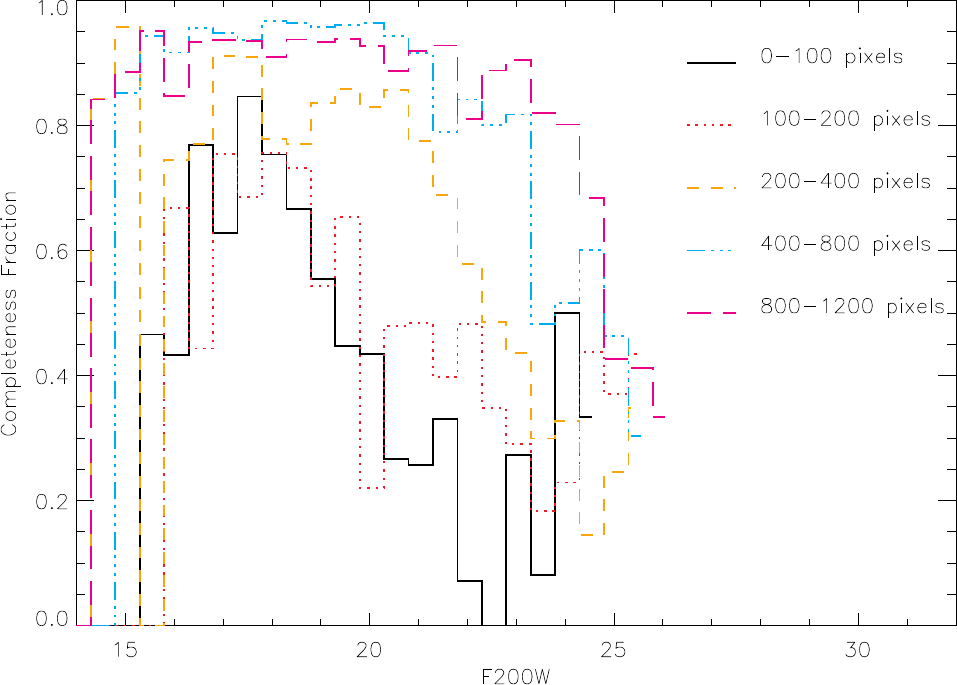}
      \caption{The fraction of recovered stars from the artificial star experiments. { The completeness is shown as a function of radius from the cluster center, illustrating the effects of crowding and nebulosity.} }
         \label{completeness}
   \end{figure}

Fig.~\ref{completeness} shows the completeness as a function of magnitude, both within the cluster and in the general field. Away from the cluster center (which is defined in Section 3), the recovery fraction is close to 100\%\ across most of the magnitude range, with a slow decline for F200W $>$ 22 mag. Closer to the cluster center the presence of nebulosity affects the completeness in several ways. The nebulosity results in a higher background and hence background noise and the fainter sources are more easily lost. The density of stars is also higher resulting in predominantly fainter sources being lost. The high background also means a lower completeness for brighter sources since the saturation limit is more easily reached. In the centre of the cluster (0.065 pc radius) the completeness is, in general, low across all magnitudes, but in particular for faint magnitudes, as expected. { The completeness falls below 60\%\ at around F200W = 19.0, corresponding to a $0.2\:M_\odot$ half-solar metallicity, 1 Myr old star seen through an extinction of $A_V=10$ \citep{baraffe15,allard}}. Outside a radius of 200 pixels (6\arcsec), the 60\%\ completeness is above F200W = 22, well into the brown dwarf regime. 

The completeness correction for each individual object is determined by using the completeness determined for the artificial stars in the same radial bin as the observed object. A Fermi function is fitted to the completeness curves \citep{gennaro,andersen_wd1}, and the corresponding completeness for the object is determined from that fit.

\begin{table}
\caption{The 60\%\ completeness limit for the observations as a function of radius from the cluster center.  }
\begin{tabular}{lccc}
Annulus & 60\%\ mag  & 60\%\ Mass ($M_\odot$) & 60\%\ Mass ($M_\odot$) \\

pixels & F200W & 1 Myr & 2 Myr \\ 
\hline
0-100  &18.88 & 0.20 & 0.40 \\
100-200  &18.76 & 0.30 & 0.40 \\
200-400  &22.15 & 0.03 & 0.04 \\
400-800  &23.39 & 0.01 & 0.02 \\
800-1200  &24.75 & 0.01 & 0.01 \\

\hline
\label{compl_table}
\end{tabular}
\tablefoot{The corresponding 60\%\ limiting mass assuming a 1 Myr and 2 Myr solar metallicity isochrone, a distance of 4.5 kpc, and adopting an extinction of $A_V=15$.}
\end{table}

The photometric depth of the annuli are determined as the magnitude where the 60\%\ completeness limit is reached \citep[as in, e.g.,][]{gennaro,andersen_wd1}. Table~\ref{compl_table} summarizes the limits as a function of radius from the center of the cluster. As expected, because of the nebulosity and surface density of the cluster members, the completeness is a strong function of radius. We note that outside 200 pixels, which corresponds to 0.13~pc for a distance of 4.5~kpc, the observations are sensitive and over 60\%\ complete { for a   $30\:M_\mathrm{Jup}$ for a 1 Myr isochrone and $40\:M_\mathrm{Jup}$ for a 2 Myr isochrone seen through an extinction of A$_\mathrm{V}=15mag$ beyond 200 pixels from the cluster center. At higher masses, less extinction, or larger radii the completeness is higher.}  We adopt the 60\%\ completeness limit in the subsequent analysis.  The photometry is corrected for the incompleteness on a star by star basis in constructing the IMF histograms. 

\section{Results and analysis}

We present direct results from source counts and photometry. The luminosity functions for the filters and both modules are discussed before the color-magnitude diagrams, and the radial profile of the cluster is presented. 

\subsection{Luminosity functions}

\begin{figure}
   \centering
   \includegraphics[angle=0,width=9cm]{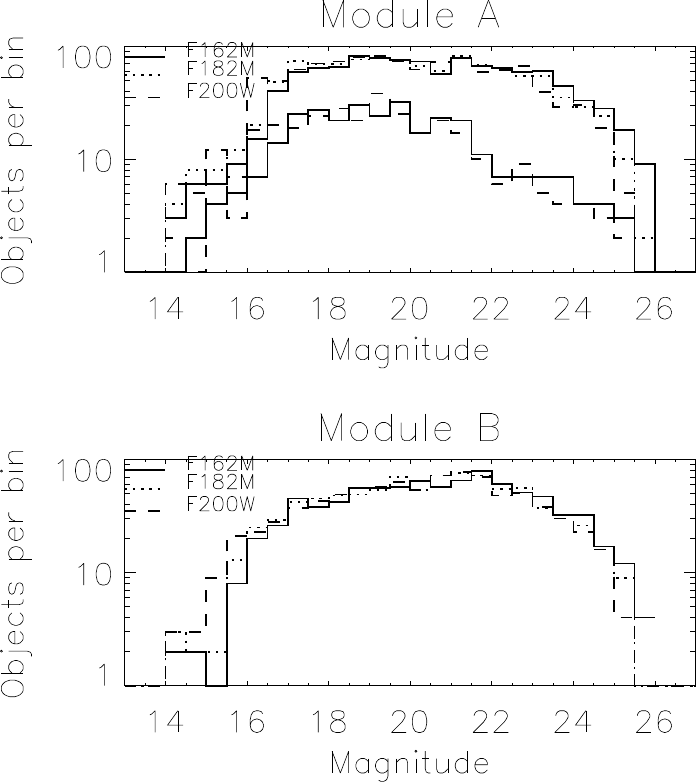}
      \caption{{ The observed luminosity functions for the \filter{F162M}, \filter{F182M}, and \filter{F200W} bands for all objects in both modules.  The lower histograms in the A module are for the central 1000 pixels around EC1.}             }
         \label{lumfuncs}
   \end{figure}

Fig.~\ref{lumfuncs} shows the uncorrected luminosity functions for the three filters and for the A and B modules. 
The shapes of the luminosity functions for the different filters are similar with a slight shift with the filter, as expected for the colors of both pre-main sequence cluster members and field stars.

The gradual turnover for both the cluster and control fields at faint magnitudes is largely due to the lack of field stars towards the antic-enter at faint magnitudes. 
This is in agreement with Galactic models \citep[e.g.,][]{robin}. 
Galaxies do contaminate the observations, but due to the depth of the observations there are relatively few galaxies and for the magnitude range of these observations the small faint galaxies are not detected \citep{rieke}. 

\subsection{Color-magnitude diagrams}

\begin{figure*}
   \centering
   \includegraphics[angle=0,width=18cm]{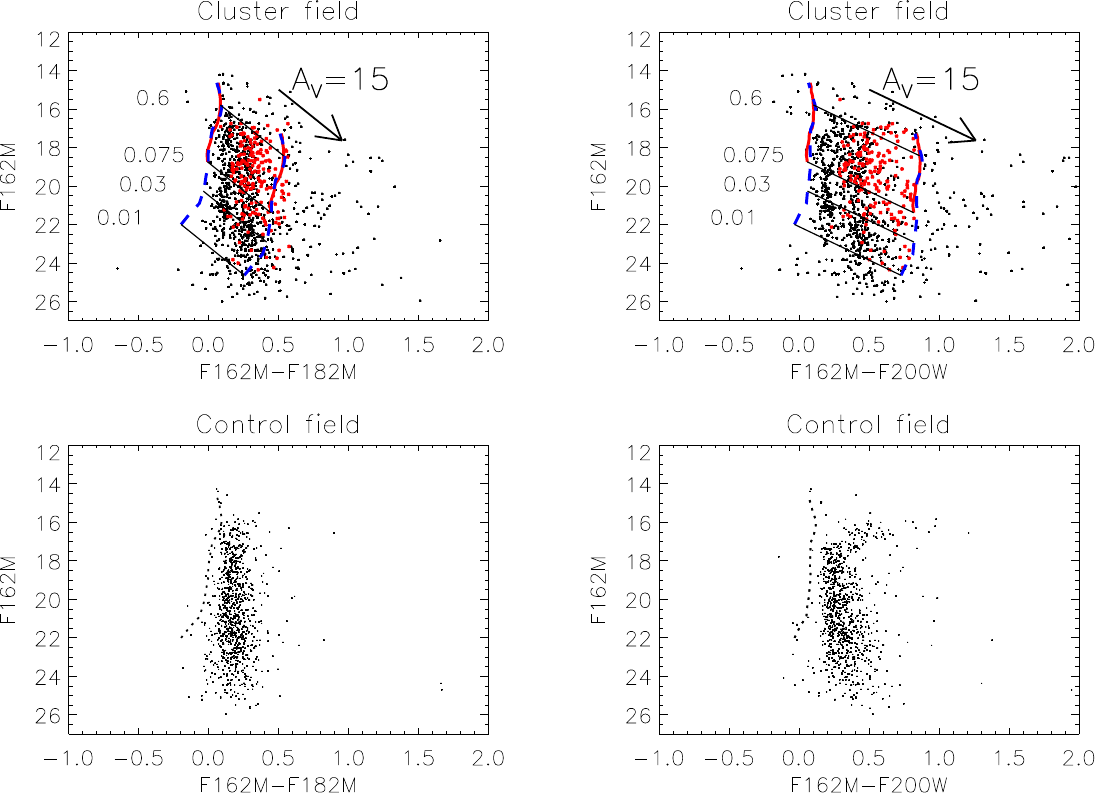}
      \caption{The CMDs for the F162M, F182M, and F200W filters. Overplotted are solar (dotted lines in blue) and half-solar 1 Myr isochrones (solid lines in red) from \cite{baraffe15} using the atmospheres from \citet{allard}. To illustrate the location of the extinction-limited samples, the isochrones have also been shifted by $A_V=15$. The locations of $0.6\:M_\odot$ and $0.075\:M_\odot$ stars are shown, the latter being the lowest mass of the half-solar metallicity isochrone. 
      The $0.6\:M_\odot$ for a 1 Myr isochrone is also the brightest object considered in the fits to the IMF. 
      The extinction vector corresponding to $A_V=15$ is shown in the color-magnitude diagrams for the cluster field. 
      Bottom panels: Similar data points as in the upper panels, but for the control region.       A 1 Myr solar metallicity isochrone, shifted to the distance of S284 is shown.} 
         \label{CMDs}
   \end{figure*}

Fig.~\ref{CMDs} shows the \filter{F162M-F182M} vs \filter{F162M} and \filter{F162M-F200W} vs \filter{F200W} color-magnitude diagrams for both modules. Overplotted are 1 Myr solar metallicity, and half solar metallicity isochrones, an extinction vector corresponding to $A_V=15$ and masses are marked on the 1 Myr solar metallicity isochrone. The location of the 1 Myr solar metallicity isochrone reddened by $A_V=15$ is also shown. Objects within 1000-pixel radius from the cluster center are marked in red. The CMDs for the control field are much more concentrated than for the cluster field, confirming the presence of extinction across the field of view of the cluster in the A module. This is more pronounced for the central region, i.e., red points, where a reddening of up to $A_V=15$ is observed. 

\begin{figure}
   \centering
   \includegraphics[angle=0,width=9cm]{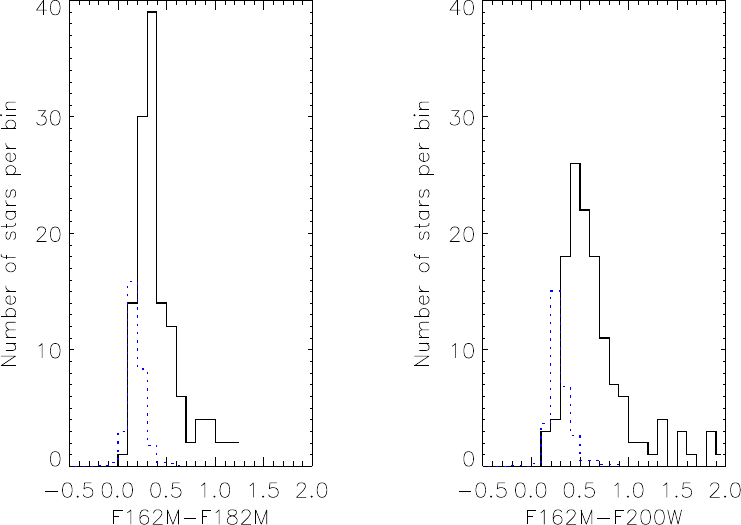}
      \caption{The distribution of colors for the cluster region and the control field scaled to the same area as the cluster region for both F162M-F182M and F162M-F200W. The control field has been scaled to the same area as the cluster circle. 
 }
         \label{color_dist}
   \end{figure}

The different distributions are also seen in the histogram of the colors between the cluster and control fields in Fig.~\ref{color_dist}. The reddening of the cluster necessitates extinction limited samples for an unbiased view of the cluster population across masses. Based on the estimated extinction values for the control field, there is a foreground extinction towards S284-EC1 of $A_V=4$, which can both be along the general line of sight and from the larger star-forming region. 

The field population, as traced by the B module is almost exclusively located at blue colors of F162M-F182M$\sim$0.3, whereas the A module region is peaked at 0.4 mag, extending to higher values. This indicates an extinction-limited sample excluding the low extinction objects will be effective in removing most of the field contamination from the cluster field (see below). 

\subsection{Cluster center and cluster radius}\label{cluster_specs}

\begin{figure}
   \centering
   \includegraphics[angle=0,width=9cm]{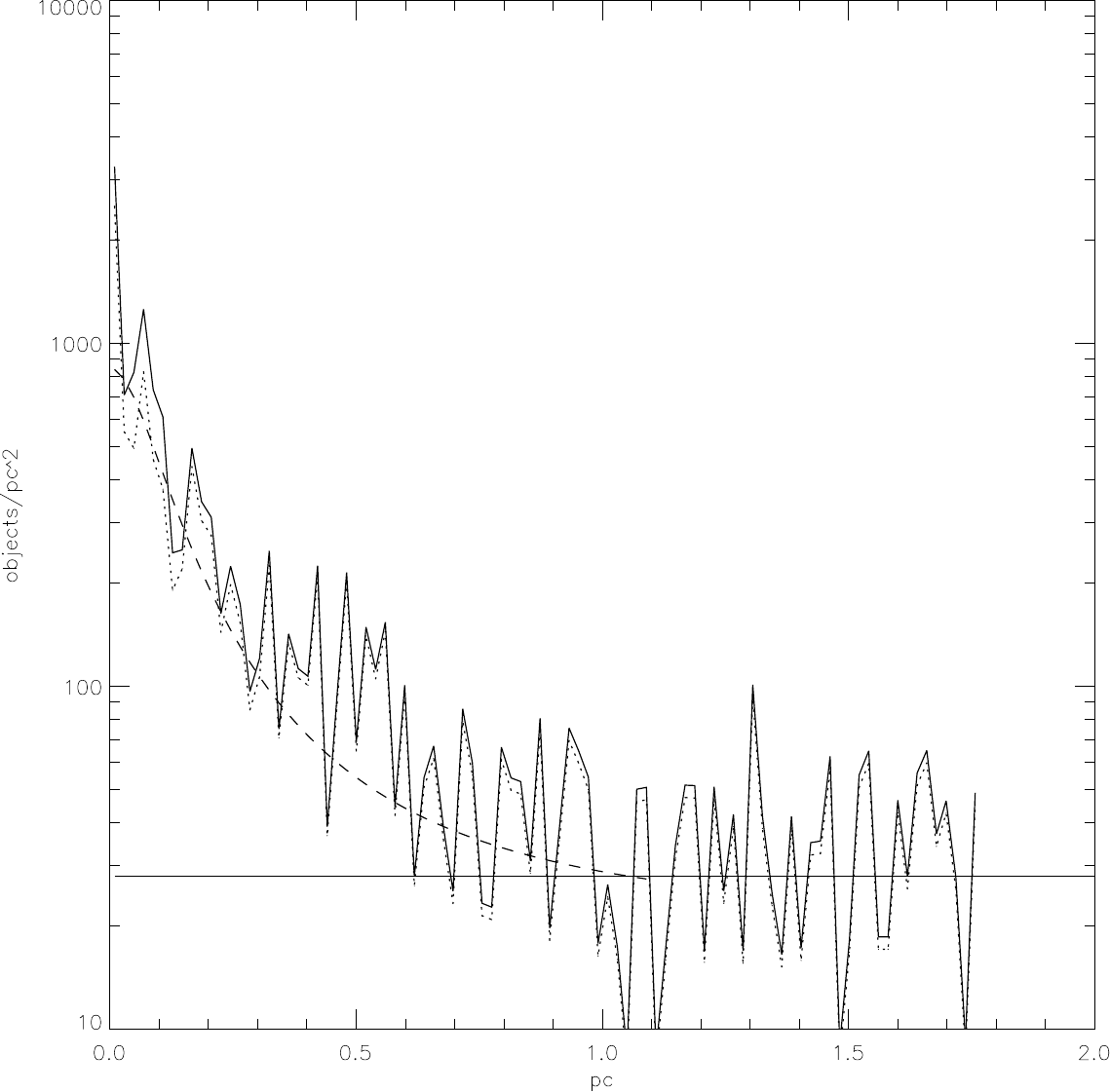}
      \caption{{ The radial profile centered on EC1, i.e.  (RA,DEC) = (6:46:15.8501,0:06:26.197) for module A. The solid horizontal line indicates the surface number density of stars in the B module, taken as the field population surface number density. Objects brighter than F200W = 19.5 are used, to ensure completeness in the central parts of the cluster as outlined above. The best-fit King profile is shown with a core radius of 0.1pc.} }
         \label{radial}
   \end{figure}

We adopt the massive protostar discussed in \citet{cheng_jwst} as the center of the cluster. The size of the cluster is determined as the radius where the surface number density determined from the radial profile is similar to that of the surface number density in the control field. The star counts are corrected for completeness. Due to the completeness limit in the central parts of the cluster, only stars down to \filter{F200W} = 19.55 are used for the profile. Figure~\ref{radial} shows the surface number density profile for the cluster. The horizontal line shows the surface number density in the control field for the same magnitude limit. 

The cluster surface number density profile follows a King profile with a core radius of 0.1~pc out to a distance of 1~pc.

We adopt a radius of 1000 pixels, corresponding to 0.65~pc, for the cluster in the following analysis. This corresponds to the point where the cluster surface number density is about 20\%\ of the background population and is a compromise between including more cluster members in the sample and limiting the background contribution. 
Within the 1000 pixel radius from the cluster center there is a total of 305 objects.

\section{The initial mass function of S284-EC1}

Here we discuss the steps involved in the estimation of the IMF.
The  isochrones are created based on stellar evolutionary models and atmospheres, with an investigation of the effects of different metallicity. The extinction and mass for the individual objects are determined de-reddening the sources to the isochrones and the adopted mass ranges suitable for determining the IMF, the creation of extinction-limited samples and the field star contamination of the cluster field are investigated. Power-law and log-normal fits are done to the sources from the extinction-limited samples and the results are compared with those from nearby star-forming regions and the Galactic field. Potential biases in the derived IMF are discussed. 

\subsection{Isochrones for low-mass objects} 

We adopt the interior models from \citet{baraffe15} for the appropriate isochrone ages and metallicity. The effective temperature and gravity from the \citet{baraffe15} models are then used as input for the \citet{allard} atmospheric models. These are then convolved with the appropriate JWST filter profiles to obtain the absolute magnitude in the observed pass bands. 

Whereas the \citet{baraffe15} models extend from $1.4\:M_\odot$ down to $0.01\:M_\odot$ for solar metallicity, they reach down to the brown dwarf limit for the $[\frac{Z}{Z_\odot}]=-0.5$ calculations (the brown dwarf limit is $0.075\:M_\odot$ for this metallicity). The metallicity of S284 has been measured to be about a third to half solar \citep{neugerella15}. The closest in metallicity from \citet{baraffe15,allard09} models are thus the $-0.5$ dex calculations.

Since the sub-solar metallicity atmospheric models do not extend below roughly the brown dwarf limit, we are forced to use the solar metallicity models to estimate the properties of the individual brown dwarfs and to estimate the full low-mass IMF in the outer parts of the cluster. This illustrates the need for future improved low-metallicity models application for low masses and young ages.

The 1 Myr solar and half-solar metallicity isochrones are plotted in Fig.~\ref{CMDs}. 
The difference in metallicity shows itself in two different ways: the low-metallicity track is bluer (hotter) and for a given mass the stars are more luminous. This is largely due to the lower opacity at lower metallicity. The combination of the low metallicity tracks being brighter and blue can to some degree be mimicked as extinction, However, the effect is relatively subtle. The difference in magnitude for a given mass for a 1 Myr and 2 Myr isochrone depends on the mass, but can be 0.1 mag. We have compared the results of fitting down { to $0.03\:M_\odot$} using the solar metallicity isochrones with the fits only extending to $0.08\:M_\odot$ using the -0.5 dex metallicity isochrones below. 

\subsection{Field object contamination}

An inherent issue in determining the cluster population is distinguishing cluster members and the general field star population. 
However, given the location of S284 towards the Galactic antic-enter from the Sun, the field star contamination is relatively minor compared to other clusters in the Milky Way. 
In particular, due to the decreasing number density of stars as a function of Galactocentric distance, the majority of field stars within the field of view will be foreground stars. 

We utilize the B module as a measure of the field star population towards S284-EC1.  
Within the B module we determine a surface number density of 120 $\mathrm{stars/arcm^2}$. 
The Galactic models by \citet{robin} predict 135 $\mathrm{stars/arcm^2}$, in agreement with what is measured in the B module within 1.5$\sigma$. The color-magnitude diagrams for the B module are thus used as the control field, as discussed below. 

The sensitivity of JWST means that galaxies are a potential contaminant within our field of view. 
Indeed, the images do contain several galaxies that are easily identified by eye. Comparing with the JADES results on the surface number density of galaxies \citep{rieke}, the contamination is rather limited for the magnitude range probed in this study. Furthermore, the brighter galaxies tend to be well resolved and are thus excluded from the photometric catalogue based on their fwhm, roundness, and sharpness. Finally, the general line of sight population is expected to be statistically the same for the two modules with different extinction distributions, and thus they will be removed as a part of the statistical field star subtraction and the extinction-limited samples. 

\subsection{Mass range and extinction-limited samples}

The mass and extinction are determined for each object by de-reddening it in the color-magnitude diagram to the adopted isochrone, following the approach in, e.g., \citet{andersen_wd1}. 
The extinction law is adopted from \citet{cardelli}, convolved with the JWST filter profiles. 
The mass of each object is thus derived using the F162M-F182M vs F162M and F162M-F200W vs F162M color-magnitude diagrams independently and assuming either a 1 or 2 Myr isochrone. 

The mass range fitted is affected by the adopted distance to the cluster, the completeness of the observations, the minimum and maximum extinction for the extinction limited-sample, and the stellar evolutionary models adopted. The IMF is fitted under different assumptions of the parameters within their expected range. Specifically, the IMF is calculated under the assumption of a distance of 4.2, 4.5, and 4.8~kpc to cover the uncertainty presented in \citet{neugerella15}, and adopting either a 1 or 2 Myr isochrone. These cover the range of ages suggested in \citet{guarcello}. However, this was determined for the cluster as a whole and the age distribution was found to be quite wide. We favor a young age for the S284-EC1 cluster based on several factors: the cluster is deeply embedded, with a typical extinction of 10$\sim$ A$_\mathrm{V}$, high considering the expected lower dust to gas ratio for a metallicity similar to the LMC  \citep{koornneef}. Further, the region is associated with strong protostellar outflows \citep{cheng_jwst,jadhav}, which further suggests youth. Thus we adopt the 1~Myr as our fiducial case, but examine the impact on the IMF results if the older age is used.

\begin{figure}
   \centering
   \includegraphics[angle=0,width=9cm]{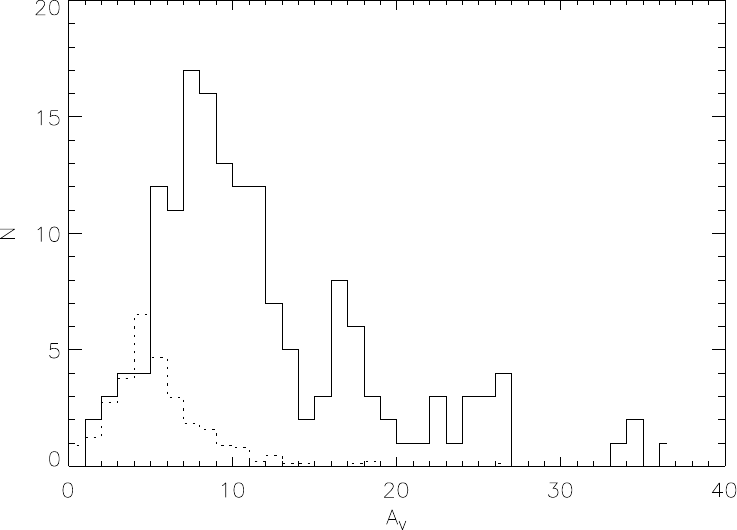}
      \caption{The distribution of extinction values for the central 1000 pixel radius of the cluster (solid histogram) and for the B module scaled to the same area as the cluster circle. A -0.5 dex, 1 Myr isochrone was assumed and objects in the mass range $0.1-0.7\:M_\odot$ have been sampled.  }
         \label{ext_dist}
   \end{figure}

Figure~\ref{ext_dist} shows the extinction distribution for the stellar content within a 1000 pixel radius of the cluster center, assuming the 1 Myr low-metallicity isochrone. The results are similar for the solar metallicity isochrone and for an age of 2 Myr. The distribution peaks at an extinction of $A_V=8$ and extends to larger values. The extinction values for field objects, as determined by projecting the objects to the -0.5 dex metallicity 1 Myr isochrone, are shown as the dotted histogram\footnote{The true extinction for the field objects should be determined knowing their spectral type and then de-reddening based on this information. However, since the purpose here is to remove the field contribution, de-reddening to the isochrone illustrates their contribution to the cluster field.}. Most of the objects have a relatively low extinction of $A_V=4.5$. Thus to limit the field star contamination we adopt a minimum extinction of $A_V=4.5$ for the cluster sample.

Since low-mass objects with high extinction would be lost due to the sensitivity limit of the observations, one can bias the IMF if an extinction-limited sample is not created \citep[e.g.,][]{andersen06}. For high extinction, we limit the sample to $A_V=15$, which, as noted in Table~\ref{compl_table}, corresponds to a mass limit of { $0.03\:M_\odot$}. A deeper extinction limit would increase the lower mass limit, while at the same time only add a few more objects at higher masses. 
 The final number of detected sources in the cluster out to a 1000 pixel radius is thus 135 in the mass interval $0.03-0.6\:M_\odot$ for a 1 Myr isochrone in the F162M-F182M color-magnitude diagram. 

\subsection{Initial mass function}
{ Building on the results in Sections 3 and 4.1-4.3 we have determined the Initial Mass Function for the cluster EC1 located East-South-East of Dolidze 25. 
Adopting a 1 Myr isochrone and a distance of 4.5 kpc the objects have been de-reddened in the color-magnitude diagrams. An extinction limited sample has been constructed based on Fig. 9 to reduce the foreground contamination. The sample used for the IMF has determined extinction values between A$_V$=4.5-15. Based on the completeness corrections summarized in Fig. 4 we determine that at 200 pixels radius (0.13 pc) the data are 60\%\ complete for a 30 Mjup object seen through A$_V$=15. For larger radii, out to 1000 pixels (0.65pc), the maximum radius we use for the sample,  the completeness is higher, resulting in a higher overall completeness. The overall completeness in the lowest mass bin is 84\%, raising to 90\%\ for higher masses.  The completeness corrected and field star subtracted Initial Mass Function is calculated and presented in Table 2. To test the sensitivity of the results we further present the results for an adapted age of 2 Myr as well as varying the distance between 4.2 and 4.8 kpc.}
\subsubsection{Log-normal fits to the IMF}

\begin{figure*}
   \centering
   \includegraphics[angle=0,width=18cm]{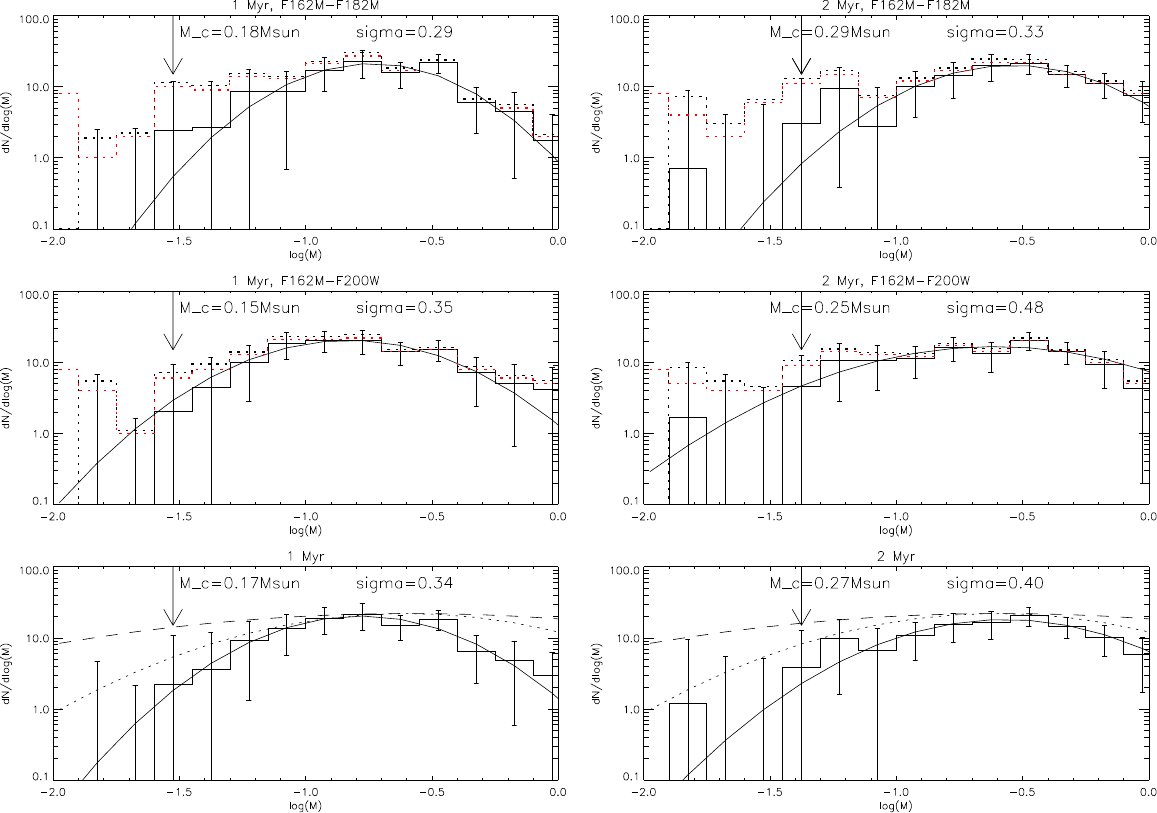}
      \caption{The IMF for the annulus 200-1000 pixels from the cluster center. Both the F162M-F182M vs F162M and F162M-F200W vs F200W color-magnitude diagrams have been used for the mass estimates of individual sources and ages of 1 and 2 Myr solar metallicity isochrones have been adopted. The limit of the fits, i.e., the 60\%\ completeness limit, is shown as an arrow in each diagram and is { $0.03\:M_\odot$ and $0.04\:M_\odot$ }for ages of 1 and 2 Myr, respectively. In each plot the dotted histogram shows the completeness corrected IMF before field subtraction and the solid lined histogram the field star corrected IMF. The red dotted lines show the histogram for the cluster field before completeness correction and field subtraction. The two bottom panels are the average of the two filter combinations. Overplotted are the \citet{chabrier05} IMF as a dotted line as well as the ONC IMF determined by \citet{gennaro_onc} as a dashed line.}
         \label{MF}
   \end{figure*}

Based on the assumptions outlined above, we have {fitted a log-normal function to the IMF for 1 and 2 Myr isochrones,  to the completeness limit of the data of 60\%  and for objects below a mass of $0.6\:M_\odot$, at which point the objects saturate. The errors are assume to the square root of the number of objects in each bin. The mass functions are in general well fitted by a log-normal distribution. Figure~\ref{MF} shows the mass functions and best log-normal fits for the extinction-limited sample, adopting a distance of 4.5 kpc, for ages of 1 and 2 Myr, for de-reddening the sources in both combinations of the color-magnitude diagrams, and for the region defined by the annulus from 200-1000 pixels in radius from cluster center. The annulus is chosen due to it being able to probe to a mass limit of  $0.03-0.04\:M_\odot$}, a substantially lower mass limit than in the center of the cluster.  

 {   
For the fiducial age of 1 Myr the best-fit log-normal distributions yield a mean peak mass of $m_c=0.16\pm0.02\:M_\odot$. 
The peaks and widths of the log-normal fits are listed in Table~\ref{mass_funcs_fits_chab}, both for the mass functions in Fig.~\ref{MF} and for the cases of different adopted distances and ages of the cluster. 

We see that our fiducial value for the peak of the IMF is significantly smaller than the peak of the field IMF of $0.25\:M_\odot$ \citep{chabrier05}, independent of the filter combination and the distance adopted. 
{ Our value is also similar to the low peak masses found in the survey by \citet{damian21} of a range of clusters within 2kpc of the Sun.} The dependency on age and distance is as expected: assuming an older or more distant population skews the IMF to higher masses. 
{  Table~\ref{mass_funcs_fits_chab} shows the fits based on adopting a shorter and larger distance. The change in the peak mass varies by $\pm$0.02 M$_\odot$. 
A pre-main sequence object gets fainter with age and thus a given observed magnitude will be interpreted as a higher mass object.}  
In particular, we note that if we adopt an age of 2 Myr the peak mass of the IMF rises to a mean value of $0.25\pm0.03\:M_\odot$.  The implications of our derived IMF parameters are discussed further in Section 4.5.}

\begin{table*}
\caption{The parameters for the log-normal distribution fits with the assumed ages and distances noted adopting the solar metallicity evolutionary models of \citet{baraffe15}. }
\begin{tabular}{cccccccccc}
Age & Filters  & Dist & $m_c$ & $\sigma$ & Filters & $m_c$ & $\sigma$  & bin\\
Myr & & kpc & $M_\odot$ & $M_\odot$ & & $M_\odot$ & $M_\odot$\\ 
\hline

1 & F162M,F182M  & 4.5 & 0.18$\pm$0.03&0.29$\pm$0.08  & F162M,F200W & 0.15$\pm$0.03&0.36$\pm$0.1 & 0.15\\
2 & F162M,F182M  & 4.5 & 0.29$\pm$0.05&0.33$\pm$0.10  & F162M,F200W &0.27$\pm$0.06&0.40$\pm$0.13 & 0.15\\
1 & F162M,F182M  & 4.2 & 0.16$\pm$0.03&0.35$\pm$0.10 & F162M,F200W  & 0.14$\pm$0.02&0.36$\pm$0.09 & 0.15\\
2 & F162M,F182M  & 4.2 & 0.22$\pm$0.03&0.34$\pm$0.08 & F162M,F200W   &0.22$\pm$0.04&0.40$\pm$0.11 & 0.15\\

1 & F162M,F182M  & 4.8 & 0.18$\pm$0.04&0.38$\pm$0.12 & F162M,F200W & 0.17$\pm$0.03&0.35$\pm$0.10 & 0.15\\
2 & F162M,F182M  & 4.8 & 0.28$\pm$0.04&0.35$\pm$0.09 & F162M,F200W  &0.28$\pm$0.04&0.33$\pm$0.08 & 0.15\\

1 & F162M,F182M  & 4.5 & 0.16$\pm$0.05&0.45$\pm$0.19& F162M,F200W  & 0.15$\pm$0.03&0.35$\pm$0.10 & 0.10 \\
2 & F162M,F182M  & 4.5 & 0.31$\pm$0.08&0.36$\pm$0.13 & F162M,F200W  &0.29$\pm$0.06&0.36$\pm$0.12 & 0.10 \\

\label{mass_funcs_fits_chab}
\end{tabular}
\end{table*}

We have performed the same fits to the stellar content, but now using the models with sub-solar metallicities. The cluster center is again avoided due to the completeness limit. The fitting is done in the same manner as for the solar metallicity models, changing age and distance in the same manner. 

\begin{table*}
\caption{The parameters for the log-normal distribution fits with the assumed ages and distances noted adopting the evolutionary models of \citet{baraffe15} with a metallicity of $-0.5$ dex. All bin are 0.1 dex.}
\begin{tabular}{cccccccccc}
Age & Filters  & Dist & $m_c$ & $\sigma$ & Filters & $m_c$ & $\sigma$ \\
Myr & & kpc & $M_\odot$ & $M_\odot$ & & $M_\odot$ & $M_\odot$\\ 
\hline

1 & F162M,F182M  & 4.5 & 0.17$\pm$0.05&0.40$\pm$0.18  & F162M,F200W  & 0.156$\pm$0.06&0.43$\pm$0.22\\
2 & F162M,F182M  & 4.5 & 0.28$\pm$0.04&0.29$\pm$0.07  & F162M,F200W &0.26$\pm$0.03&0.27$\pm$0.07\\

1 & F162M,F182M  & 4.2 & 0.16$\pm$0.07&0.44$\pm$0.23 & F162M,F200W  & 0.10$\pm$0.11&0.55$\pm$0.44\\

2 & F162M,F182M  & 4.2 & 0.24$\pm$0.03&0.30$\pm$0.07  & F162M,F200W   &0.24$\pm$0.03&0.31$\pm$0.08\\

1 & F162M,F182M  & 4.8 & 0.21$\pm$0.04&0.29$\pm$0.09 & F162M,F200W  & 0.12$\pm$0.20&0.74$\pm$1.11\\

2 & F162M,F182M  & 4.8 & 0.28$\pm$0.05&0.32$\pm$0.10  & F162M,F200W  &0.26$\pm$0.04&0.30$\pm$0.07\\

\label{mass_funcs_fits_chab_half}
\end{tabular}
\end{table*}

The fits provide results that are broadly consistent with those obtained with the solar metallicity fits. In particular, when adopting a 1 Myr isochrone the peak mass is found to be about $0.16\pm0.05\:M_\odot$, which is significantly below that of the local Galactic field population. However, again, if a 2 Myr isochrone is adopted, this mass scale is raised and would be consistent with that of the local value.
We note the uncertainties in the fit results for the lower-$Z$ models are larger due to the smaller fit range used.

\subsubsection{Cluster-wide power-law fits to the stellar content}

The log-normal fits were carried out for the regions of the cluster with the highest completeness, i.e., outside 200 pixel radius. However, despite the poor completeness for both the brightest and faintest stars closer to the center, the IMF here can still be probed over a more limited mass range. Table~\ref{fit_powlaw} summarizes the best-fit power-laws for the mass range $0.2-0.7\:M_\odot$ for 1 Myr isochrones and $0.3-0.8\:M_\odot$ for the 2 Myr isochrones for the whole cluster, i.e., from 0-1000 pixels in radius.  

\begin{table}
\caption{The best-fit power laws to the whole cluster region over the mass range $0.2-0.6\:M_\odot$ for a 1 Myr isochrone and $0.3-0.7\:M_\odot$ for a 2 Myr isochrone. }
\begin{tabular}{ccccc}
Age & Filters  & Dist & $log(\frac{Z}{Z_\odot})$ & $\Gamma$ \\
Myr & & kpc \\
\hline
1 & F162M,F182M  & 4.5 & 0&  -1.0$\pm$0.54\\
2 & F162M,F182M  & 4.5 & 0& -1.4$\pm$0.56\\
1 & F162M,F200W  & 4.5 & 0& -1.3$\pm$0.73\\
2 & F162M,F200W  & 4.5 & 0& -0.8$\pm$0.66\\
1 & F162M,F182M  & 4.5 & -0.5 & -1.0$\pm$0.54\\
2 & F162M,F182M  & 4.5 & -0.5 & -1.0$\pm$0.33\\
1 & F162M,F200W  & 4.5 & -0.5 & -1.3$\pm$0.73\\
2 & F162M,F200W  & 4.5 & -0.5 & -0.8$\pm$0.66\\

\hline
\label{fit_powlaw}
\end{tabular}
\end{table}

The uncertainties on the individual power-law fits are relatively large due to the small mass range fitted, dictated by the completeness limit and the saturation limit. Since the extinction cannot be estimated for the higher mass stars that are saturated,
they are not included in the discussion. 
The average power law index for the different fits and metallicities is $-1.0\pm0.3$, i.e., slightly shallower, but still consistent with, the Salpeter {value, and also consistent with the slightly top-heavy higher mass IMFs measured in the LMC \citep{schneider} and in other outer Galaxy star-forming regions \citep{yasui23}.}

\subsection{IMF biases}

{
Several factors can affect the derived IMFs in addition to age and distance. 
These include circumstellar disks, binarity, choice of extinction law, and the metallicity used to create the evolutionary tracks, which we discuss below.

Circumstellar disks are common around YSOs. The inner part of the disk can be heated to 2000~K and they can in some cases be detected through their infrared emission at and beyond 2 $\mu$m, i.e., excess to the stellar photosphere. If not corrected for, this excess emission could bias the determined IMF. \citet{muzic19} performed a detailed analysis of the effects of disks on the IMF for the Rosette nebula, which has an intermediate disk fraction. They found the effects on the determined masses were typically minor and thus the IMF was only marginally affected. Nevertheless, we have investigated the effect assuming a circumstellar disk fraction of 40\%\ and a disk excess in the F200W band uniformly chosen between 0 and 0.2 mag. For the 1 Myr isochrone the change in the peak mass is modest, increasing the peak mass by $0.01\:M_\odot$, confirming the results from \citet{muzic19} that the impact of circumstellar disks on the IMF is very limited.

There are a variety of extinction laws that may be applicable to dust in the diffuse ISM and denser star-forming clouds.
To illustrate the difference for the IMF that is induced using the extinction law of \citet{fitzpatrick} rather than that of \citet{cardelli}, we see that the best log-normal fit using the F162M-F182M vs F162M color magnitude diagram and the 1 Myr isochrone is 0.18 M$_\odot$, and $\sigma=0.34$. 
Thus, the impact of using this different extinction law is very minor for the range of extinction values in this study.

The field IMF peaks at 0.25~$M_\odot$ for the system IMF. Thus, if the binary properties would be different at low-metallicity this could affect the comparison. However, even the single-object IMF in \citet{chabrier05} only sees a shift in the peak mass to $0.2\:M_\odot$, marginally consistent with the highest peak masses we determine. Furthermore, \citet{machida} suggests from theory that the binary fraction should be higher in a low-metallicity star forming region, which would only lower the deduced single object IMF from the fits in Table~\ref{mass_funcs_fits_chab}. 

The IMFs and thus peak masses could be biased by the usage of the solar metallicity isochrones. To test for this we have fitted log-normal distributions to the stellar range up to the saturation limit using the sub-solar metallicity isochrones. 
For a distance of 4.5 kpc we determined a peak mass of $0.13\pm0.09\:M_\odot$ for 1 Myr and $0.26 M_\odot$ for the 2 Myr isochrones. 
The peak is about 10\% higher for the 2 Myr isochrone compared to the peak for solar metallicity, but the conclusion remains the same. The 1 Myr isochrone provides a peak value consistent with that of the solar metallicity fits, but with larger uncertainties, mainly due to the more limited mass range and the peak being close to the lower mass limit. 
}

\subsection{Comparison with other IMF determinations} 

The local Galactic field population has been found to have a mass function that peaks at $0.25\:M_\odot$ \citep{chabrier05}. Previous studies of nearby regions with solar metallicity have found results consistent with this \citep[e.g.,][using a sample of clusters studied spectroscopically]{andersen_08}. 
{ However, there has been observed some variations from region to region \citep[e.g.,][]{bastian10}. For example, \citet{damian21} used a photometric sample of 9 regions with variable depth in terms of limiting magnitude to determine an average peak mass of $0.32\pm0.02\:M_\odot$ with an observed  range from $0.18-0.44\:M_\odot$. 
Overall, the peak of the IMF measured here is on the lower side of the values determined for solar metallicity regions.}

In contrast, recent studies of the low-mass IMF in Galactic metal-poor environments have suggested a lower peak mass. The Digel Cloud 2 has a lower metallicity than S284, i.e., $-0.7$ dex. This region was imaged by \citet{yasui24}, who determined a peak mass ${\rm log}(m_0)=-1.5\pm0.5$ ($0.03^{+0.07}_{-0.03} M_\odot$). Their 
result, would thus suggest that the peak mass of the IMF decreases relatively rapidly with decreasing metallicity. 
However, it should be noted that the isochrones used are different than the ones used here and in, e.g., \citet{gennaro_onc} and the effects of crowding and nebulosity were gauged purely by photometric errors and not completeness calculations.  Similarly, \citet{yasui23} suggest for the Galactic metal-poor region Sh2-209 that the IMF has a break mass at $0.1\:M_\odot$, but with the precise value depending on the distance and age for the system. The equivalent power law fitting to the local Galactic field population finds a break mass of $0.5\:M_\odot$ \citep{kroupa2024}.

Previous studies in, e.g., the SMC and LMC have been limited to the higher mass part of the IMF, where the power-law index was found to be similar to that of the Galactic field IMF \citep{andersen_30dor,dario}. However, the field IMF in the dwarf galaxy Coma Berenices has suggested a higher peak mass of the IMF, despite the low metallicity of the galaxy \citep{gennaro_dwarf}. Note that the completeness limit in these studies was estimated at $0.23\:M_\odot$, barely reaching the Chabrier IMF peak. As noted in \citet{andersen_wd1}, when only one side of the log-normal is fitted, there is some degeneracy between the peak mass and the width of the fit. 

Thus, in summary, a picture is emerging of a possible change in the low-mass IMF towards lower peak masses for low-metallicity environments. This trend is in qualitative agreement with the simulations of \citet{chon2021}, although their simulations did not include magnetic fields or feedback mechanisms which can limit fragmentation. In contrast, models such as that of \citet{Guszejnov2022} are more difficult to reconcile with this observational result.

The width of the best-fit distributions appear, in general, to be more narrow than that of the field IMF. A similar effect has been observed in some solar-metallicity regions, e.g., Westerlund 1 \citep{andersen_wd1}. However, in this case, the more narrow IMF is measured down to { $0.03\:M_\odot$, compared to $0.2\:M_\odot$ in Westerlund 1. }

The advent of deep, high spatial resolution near-infrared imaging with JWST now allows us to probe much larger ranges of parameter space to quantify variations in the IMF as a function of environment. With this ability to probe the IMF across the Galaxy deep into the brown dwarf regime, improved models are needed to compare with the observations, both in terms of brown dwarf and stellar evolutionary models and simulations including the relevant physics with sufficient dynamical range to cover the low-mass IMF in complex environments. 

\subsection{Cluster mass}

We can estimate the total mass of the cluster based on the measured part of the IMF. 
We have adopted the log-normal results derived from the extinction limited sample and extrapolated to higher masses using a combined mass function consisting of the log-normal fit below $1\:M_\odot$ and a Salpeter power law mass function above this value. \citet{guarcello} found for Dolidze 25 as a whole that the IMF above $1\:M_\odot$ was consistent with having a Salpeter index, similar to results found in other low-metallicity environments, e.g., in the Magellanic clouds. 

The completeness excludes reaching the lowest masses in the central region and thus we have to extrapolate the mass function there. This is done based on the radial profile in Fig.~\ref{radial}, i.e., for stars over a mass of $0.2\:M_\odot$. We have compared the derived IMF in two radial bins to probe if any mass segregation might be evident that could skew the mass determination. For the 1 Myr F162M-F182M sample at a distance of 4.5 kpc we determined the IMF in the annuli 0.13-0.46~pc and 0.46-0.66~pc. 
The peak masses are found to be $0.19\pm0.04\:M_\odot$, and $0.15\pm0.05\:M_\odot$, respectively, supporting an assumption that there is no mass segregation in the cluster. 

The IMF was determined for the extinction limited samples as discussed above. However, to obtain a better estimate of the total mass we use all objects with an extinction above the lower extinction limit of 4.5 $A_V$. The total mass observed between the lower and upper mass limit for the IMF determination is then scaled to the full IMF as described above. Finally, it is also scaled to include the mass within 0.13 pc using the radial profile fit. For an assumed ago of 1 Myr the estimated total mass is $160\:M_\odot$. If a 2 Myr age is assumed, the total mass is estimated to be $330\:M_\odot$. 
{ This mass is substantially lower than the gas mass estimated in \citet{cheng_jwst} showing the system is gas dominated confirming its youth.}

 \section{Conclusions}
 
We have presented JWST NIRCam F162M, F182M, and F200W imaging of the embedded cluster S284-EC1 that is forming in a low metallicity environment in the outer Galaxy. Our main conclusions are as follows:

\begin{enumerate}
\item The NIRCam images are sensitive to young stars and brown dwarfs between $0.6\:M_\odot$ (the saturation limit) down to { $0.03\:M_\odot$} (assuming a fiducial 1 Myr age and for an extinction of $A_V=15$). The data are at least 60\%\ complete to this mass limit outside a radius of 0.13~pc from the cluster center. 

\item Log-normal fits to the IMF of the YSO population defined from an extinction-limited sample find the peak mass to be $m_0=0.16\pm0.02\:M_\odot$ This indicates a potential dependence of IMF properties on metallicity, which is consistent with an observational trend found in lower metallicity region Digel 2 \citep{yasui24}.
\item The main caveat associated with our result for the IMF peak mass is the dependence of the result on the assumed age of the population. However, a value of 1~Myr appears to be the most reasonable, given the embedded and star-forming nature of S284-EC1.
\item The overall current stellar mass of S284-EC1 is about $160\:M_\odot$ within a radius of 0.65~pc. There is no evidence for mass segregation within the cluster in the low-mass YSO population, although at least one massive star, with current $m_*\sim 10\:M_\odot$, appears to be forming in the cluster center \citep{cheng_jwst}.

\end{enumerate}

\begin{acknowledgements}
This work is based [in part] on observations made with the NASA/ESA/CSA James Webb Space Telescope. The data were obtained from the Mikulski Archive for Space Telescopes at the Space Telescope Science Institute, which is operated by the Association of Universities for Research in Astronomy, Inc., under NASA contract NAS 5-03127 for JWST. These observations are associated with program 2317. AB acknowledges support from the Chalmers Astrophysics and Space Sciences Summer research program (CASSUM). JCT acknowledges support from ERC Advanced Grant 788829 (MSTAR).
RF acknowledges financial support from the Severo Ochoa grant CEX2021-001131-S MICIU/AEI/ 10.13039/501100011033 and PID2023-146295NB-I00.
\end{acknowledgements}

\end{document}